\documentclass[sigconf]{acmart}

\renewcommand\footnotetextcopyrightpermission[1]{} % removes footnote with conference information in first column
\usepackage{multirow}
\usepackage{booktabs}
\usepackage{xcolor}
\usepackage{blindtext}
\usepackage{enumitem}
\newcommand{\subscript}[2]{$#1  #2$}
 
\definecolor{codegreen}{rgb}{0,0.6,0}
\definecolor{codegray}{rgb}{0.5,0.5,0.5}
\definecolor{codepurple}{rgb}{0.58,0,0.82}
\definecolor{backcolour}{rgb}{0.95,0.95,0.92}

\definecolor{mygray}{gray}{0.7}

\usepackage{listings}

\lstdefinestyle{mystyle}{
    commentstyle=\color{codegreen},
    numberstyle=\tiny\color{codegray},
    stringstyle=\color{codepurple},
    basicstyle=\ttfamily\footnotesize,
    breakatwhitespace=false,         
    breaklines=true,                 
    captionpos=b,                    
    keepspaces=false,                 
    numbersep=5pt,                  
    showspaces=false,                
    showstringspaces=false,
    showtabs=false,                  
    tabsize=1
}
 
\AtBeginDocument{%
  \providecommand\BibTeX{{%
    \normalfont B\kern-0.5em{\scshape i\kern-0.25em b}\kern-0.8em\TeX}}}

\begin{document}
\fancyhead{}

\title{SATURN}
\subtitle{Software Deobfuscation Framework Based on LLVM}

\author{Peter Garba}
\authornotemark[1]
\affiliation{%
  \institution{Thales, DIS - Cybersecurity}
  \city{Munich}
  \country{Germany}
}
\email{peter.garba@thalesgroup.com}

\author{Matteo Favaro}
\affiliation{%
  \institution{Zimperium, Mobile Security}
  \city{Noale}
  \country{Italy}
}
\email{matteo.favaro@reversing.software}

\begin{abstract}
  The strength of obfuscated software has increased over the recent years. Compiler based obfuscation has become the de facto standard in the industry and recent papers also show that injection of obfuscation techniques is done at the compiler level. In this paper we discuss a generic approach for deobfuscation and recompilation of obfuscated code based on the compiler framework \textit{LLVM}. We show how binary code can be lifted back into the compiler intermediate language \textit{LLVM-IR} and explain how we recover the control flow graph of an obfuscated binary function with an iterative control flow graph construction algorithm \cite{biondi:hal-01241356} based on compiler optimizations and satisfiability modulo theories (\textbf{SMT}) solving. Our approach does not make any assumptions about the obfuscated code, but instead uses strong compiler optimizations available in \textit{LLVM} and \textit{Souper Optimizer} to simplify away the obfuscation. Our experimental results show that this approach can be effective to weaken or even remove the applied obfuscation techniques like constant unfolding, certain arithmetic--based opaque expressions, dead code insertions, bogus control flow or integer encoding found in public and commercial obfuscators. The recovered \textit{LLVM-IR} can be further processed by custom deobfuscation passes that are now applied at the same level as the injected obfuscation techniques or recompiled with one of the available \textit{LLVM} backends. The presented work is implemented in a deobfuscation tool called \textit{SATURN} (Figure~\ref{fig:SATURN}).
\end{abstract}


\begin{teaserfigure}
  \begin{center}
  \includegraphics[width=25mm]{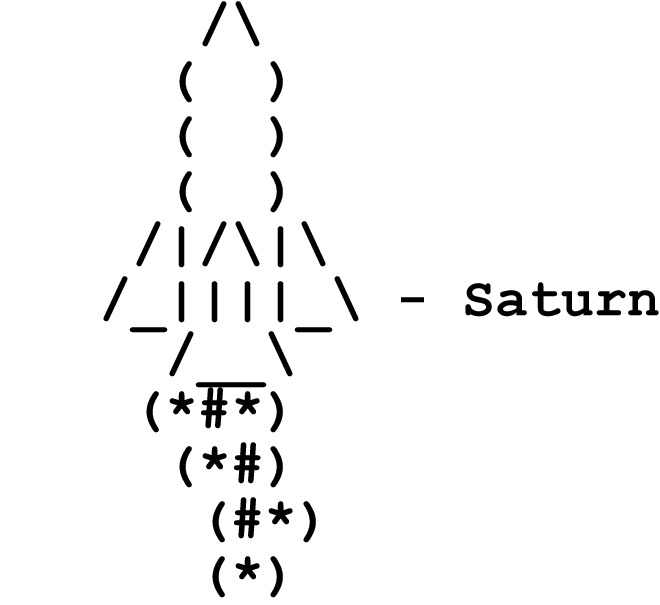}
  \label{fig:teaser}
  \end{center}
\end{teaserfigure}

\maketitle
\section{Introduction}

% A picture describing the SATURN workflow
%\begin{figure*}[hbt!]
\begin{figure}[hbt!]
 \center
  \includegraphics[width=87mm]{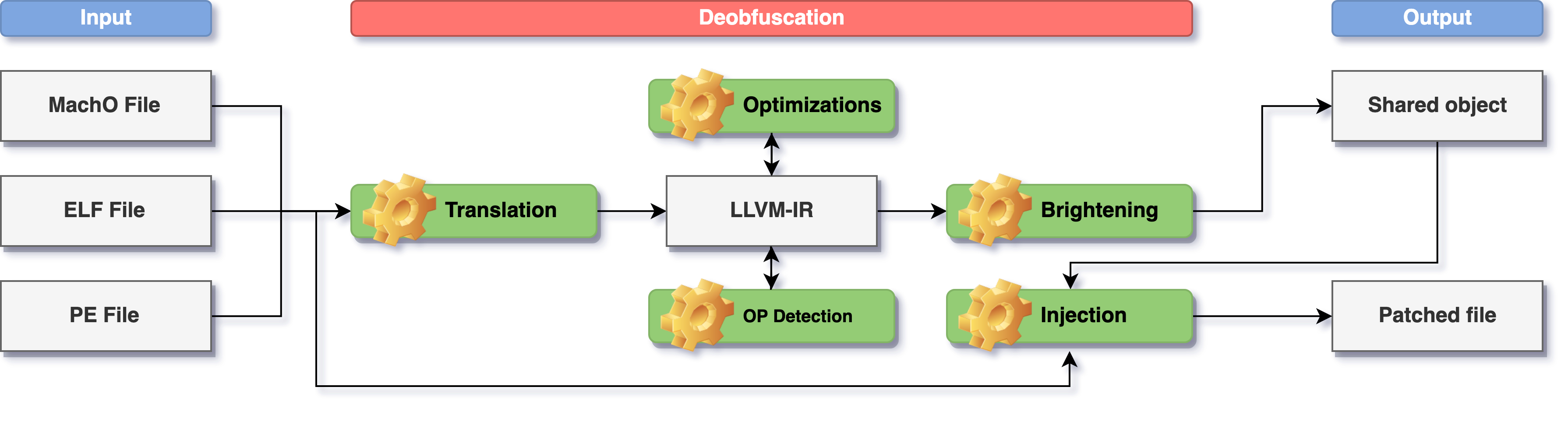}
  \caption{Workflow of the SATURN deobfuscation framework}
  \label{fig:SATURN}
%\end{figure*}
\end{figure}

%% NEW
%% TODO insert cite for ollvm, epona and
In recent years we observed the increase in popularity and rise of intermediate language and source--based obfuscators, specifically due to the growth and diverse landscape of target architectures, especially on the mobile market \cite{ieeespro2015-JunodRWM}. While classical binary--based obfuscations were previously attacked by applying pattern--based rules or simple static analysis, higher--level obfuscations applied on intermediate language or source code cannot be effectively compromised. Modern protection tools are mostly based on \textit{state--of--the--art} compiler frameworks like \textit{LLVM} that allow much more complex obfuscation logic \cite{ieeespro2015-JunodRWM} \cite{EPONA}.\\

In this paper we present an automatic deobfuscation approach based on \textit{LLVM}'s strong code optimizations. The paper focuses on several aspects that need to be addressed during deobfuscation: \textit{translation of binary code to LLVM-IR}, \textit{control flow graph recovery}, \textit{detection of opaque predicates}, \textit{deobfuscation}, \textit{brightening of the recovered function} and \textit{recompilation}.\\

% translation of binary code to LLVM-IR
Translating binary code into \textit{LLVM-IR} is not a straightforward task. A binary opcode does not only execute the operation itself, but might also address several other operations like the calculation of the condition codes/flags that influence later branch instructions. The information that could be used to translate the binary code into an intermediate language like the \textit{LLVM-IR} is normally lost during compilation and, especially in obfuscated binary code, this task can be even harder. One approach to target the problem is to implement the exact semantic of each binary opcode and store the output into a structure that holds the current state of the registers. This is a generic approach that lifts the binary code into a virtualized context but does not make any assumptions about the binary code itself. The recovered \textit{LLVM-IR} is fully functional but the readability of the IR might be very low. In this paper we make use of \textit{Remill} \cite{Remill} \cite{Korencik2019thesis} to address the problem of binary code translation.\\ 

% control flow recovery
Control flow obfuscation is a technique to hide the original control flow of a function. To deobfuscate the function the attacker has to recover the control flow graph from the obfuscated binary code. Modern obfuscation tools that operate on intermediate languages like \textit{LLVM-IR} have the ability to heavily obfuscate the control flow graph. We introduce an algorithm that makes use of the \textbf{State} struct in \textit{Remill} to recover the edges of each lifted basic block. The lifted basic blocks and edges represent the recovered control flow graph. The recovery of the control flow graph is done statically and automatically during the lifting of the obfuscated binary code. Compared to previous work (\cite{cav08}, \cite{AutomaticDeobfuscation}, \cite{vmcai12}, \cite{Wang:2015:RD:2831143.2831183}, \cite{Reinbacher:2011:PCF:2038642.2038662}) that was done on control flow graph recovery, our approach does not need any prior knowledge about the binary code and doesn't rely on traces of the function. Instead, the path exploration is done based on the partially deobfuscated basic blocks and their predecessors. Our algorithm is similar to the \textit{Iterative Control Flow Graph Construction} in \cite{biondi:hal-01241356} but is superior in the way that it works independently of the order in which the branches are examined.\\

% detection of opaque predicates
A technique to conceal the control flow graph of a function is the insertion of opaque predicates (\textbf{OP}) to thwart na\"ive control flow graph reconstruction algorithms. An opaque predicate is a conditional branch injected into the control flow graph whose condition exists to confuse or thwart reverse engineering, but whose evaluation is deterministic, and thus irrelevant to the greater logic of the program \cite{Collberg97ataxonomy}. We present an effective approach to detect and remove opaque predicates. The shown approach is based on strong \textit{LLVM} and \textit{Souper Optimizer} optimizations. For opaque predicates that are resistant to the applied optimizations and/or to verify the optimization results, we use an approach based on \textit{SMT} solving. The way to identify opaque predicates with \textit{SMT} solving is not new \cite{Ming:2015:LLO:2810103.2813617}, but we believe that the way we combine several tools and algorithms are a rich contribution to this paper.

% introduce Brightening
\begin{description}[align=left]
\item [\textbf{\textit{Brightening} [COMP.]} \textcolor{mygray}{\textit{verb}} -- ] \textit{Reshaping code to make it more readable and understandable for humans}
\end{description}

% deobfuscation
Constant unfolding, arithmetic-based opaque expressions, dead code, bogus control flow and integer encoding are not only found in hardened code, but can also appear in non-obfuscated code. Normally, during compilation of the source code, the compiler detects this kind of patterns and optimizes them away to obtain the best possible result. The presented approach relies on the reshaping of the \textit{LLVM-IR}, as the way the code gets lifted by \textit{Remill} might hinder the optimizer to reach the best result. The needed steps to reshape the \textit{LLVM-IR} are generic and don't rely on any prior knowledge about the obfuscator.\\

Without brightening, the \textit{LLVM-IR} would be fully functional but in this paper we aim to reach a \textit{vanilla}\footnote{As close as possible to a non-obfuscated compiled source code} state representation of the lifted function. This includes reconstruction of the original function arguments and transformation of the \textit{Remill} specific lifted function based on the \textbf{State} struct (Listing~\ref{RemillStruct}) into a clean \textit{LLVM} function with its original signature. \\

% Figure for Remill Struct
\begin{figure}[hbt!]
\centering
\begin{lstlisting}[language=C, 
caption=Remill \textbf{State} struct definition for x86\_64, label={RemillStruct}
]
struct State {  
  VectorReg vec[kNumVecRegisters];
  ArithFlags aflag;
  Flags rflag;
  Segments seg;
  AddressSpace addr;
  GPR gpr;
  X87Stack st;
  MMX mmx;
  FPUStatusFlags sw;
  XCR0 xcr0;
  FPU x87;
  SegmentCaches seg_caches;
}
\end{lstlisting}
\end{figure}

% Recompilation
Once the control flow graph is recovered and the function is deobfuscated, one of the goals of the presented approach is to recompile and execute the lifted function. Due to the choice of \textit{LLVM-IR} as destination language for the lifted binary code, we can easily compile the recovered code back into binary code by using one of the available \textit{LLVM} backends (X86, ARM, AArch64, RISC-V and others). \\

% Experiments
Our experiments show that we are able to apply our approach on current \textit{state-of-the-art} obfuscations and also, to partially defeat the \textit{anti-symbolic deobfuscation} tricks introduced in \cite{KillSym}.\\

Our work is not only useful for deobfuscation. In fact this approach can also be used for further applications like fuzzing, as input for dynamic symbolic execution (\textbf{DSE}) engines like \textit{KLEE} \cite{KLEE}, as input for \textit{LLVM} based obfuscators like \textit{O-LLVM} \cite{ieeespro2015-JunodRWM}, to achieve automatic payload creation for exploitation as shown in \cite{SMTRolf} or in general to recompile binary code with the best available \textit{CPU} optimizations (-march=\textbf{native}) to improve the performance of applications or to introduce new compiler based security features. This applies especially to applications where the source code is not available.\\
% End introduction

% Goals and Challenges
\subsection{Goals and Challenges}

We want to propose a deobfuscation framework based on \textit{LLVM} and its strong optimizations for real world applications. Using \textit{LLVM} for reverse engineering might look like an overcomplication in the beginning, but it's similar to what is done during the compilation of source code. The \textit{LLVM} compiler framework has all the needed tools to easily create and modify the control flow graph, its basic blocks and instructions. The challenge is to lift the binary code into the \textit{LLVM-IR} and get it into a shape that is equal to a non-obfuscated compiled source code. The techniques to reach this goal should be generic, non error-prone and lightweight. The framework should always generate working \textit{LLVM-IR} that can be recompiled and executed. We aim at proposing a framework to lift the binary code back into a clean and understandable \textit{LLVM-IR} that is built on mature tools around the \textit{LLVM} ecosystem. Our vision is to get the attack surface back to the level it was implemented at -- the compiler level.\\

% Contribution
\subsection{Contribution}

We summarize our contributions as follows.
\begin{itemize}
  \item We propose an automatic deobfuscation tool that is generic enough to deal with several obfuscation techniques.
  \item We propose a framework that can recompile and inject the \textit{LLVM-IR} code back into the given binary.
  \item We propose an effective and efficient method to identify opaque predicates at the \textit{LLVM-IR} level, that are then solved and verified using compiler optimizations and \textit{SMT} solvers.
  \item We propose a generic method to transform the binary code lifted by \textit{Remill} into a cleaner \textit{LLVM-IR} without the \textit{Remill} \textbf{State} struct. This includes the recovery of the stack and the function arguments.
  \item We show that our work can be used to weaken or even remove anti-symbolic execution tricks like the ones introduced in the work of \cite{KillSym} and allows the usage of \textit{state-of-the-art} source-level dynamic symbolic execution tools.
  \item We propose a framework that can generate a compact representation of the obfuscated constraints that are easier to solve or check for satisfiability.
  \end{itemize}

% Discussion
\subsection{Discussion}

We explore several steps to recover the binary code from an obfuscated binary, based on the lifting of the code to the compiler intermediate language \textit{LLVM-IR}. We propose several algorithms that were implemented in the tool \textit{SATURN} that help to handle different aspects of binary code deobfuscation. To our knowledge the implementation of \textit{SATURN} is \textit{state-of-the-art} and lifts the attacking surface from the obfuscated binary code back to the compiler level. Our work has a high impact on the security of obfuscated binaries and allows the usage of efficient \textit{state-of-the-art} source-/IR-level dynamic symbolic execution tools like \textit{KLEE} to further analyze the recovered code. We provide an experimental setup that includes several corner cases that might hinder binary code lifting, but also apply our method to strong obfuscated real world binaries.\\

% Background
\section{Background}
\subsection{LLVM}
"\textit{LLVM} began as a research project at the University of Illinois, with the goal of providing a modern, SSA-based compilation strategy \cite{Torczon:2007:EC:1526330} capable of supporting both static and dynamic compilation of arbitrary programming languages. Since then, LLVM has grown to be an umbrella project consisting of a number of sub-projects, many of which are being used in production by a wide variety of commercial and open source projects as well as being widely used in academic research" \cite{LLVM}. To understand our approach it's not crucial to understand how \textit{LLVM} and its internal language \textit{LLVM-IR} are designed, but the reader should keep in mind that the \textit{LLVM-IR} is based on the Static Single Assignment form (\textbf{SSA}) \cite{SSA} which makes it easier to construct the final formula passed to the SMT solver \cite{SMTRolf}. 

\subsection{Remill}
"\textit{Remill} is a static binary translator that translates machine code instructions into \textit{LLVM} bitcode. It translates x86 and amd64 machine code (including AVX and AVX512) into the \textit{LLVM-IR}" \cite{Remill}. In our work we make extensive use of \textit{Remill} to lift the binary code into the \textit{LLVM-IR}. \textit{Remill} does not make any assumptions about the stack or the arguments of a lifted function since it only lifts single instructions.\\ 

\subsection{Souper optimizer}
\textit{Souper} is particularly convenient because it's an \textit{LLVM}-based project that, with the help of \textit{KLEE} \cite{KLEE}, is capable of converting a sequence of \textit{LLVM-IR} instructions into an \textit{SMT} formula and use several \textit{SMT} solvers to discover additional peephole optimizations. As a desired side-effect we can benefit from its results to determine the opaqueness of a conditional branch. \textit{Souper} has the possibility to cache the \textit{SMT} queries and results into an external \textit{Redis} database \cite{Redis} to improve the performance. This leaves us with a database full of opaque predicates and obfuscation patterns that could be analyzed in further studies.\\

\subsection{KLEE}
"\textit{KLEE} is a symbolic execution tool capable of automatically generating tests that achieve high coverage on a diverse set of complex and environmentally-intensive programs and operates on the \textit{LLVM-IR}" \cite{KLEE}. \textit{KLEE} is not only a useful tool for testing software but it's also very effective in attacking several code obfuscation techniques. Current work proposed in \cite{KillSym} tries to hinder symbolic execution tools like \textit{KLEE} to do their work effectively.

% Motivation
\section{Motivation}
\subsection {Attacker model}
\textbf{Goal.} We consider a man-at-the-end (\textbf{MATE}) scenario where the attacker has full access to a protected binary under attack and no access to the source code or unprotected binary. The attack model and the methodology follow closely the survey by Schrittweiser et al. \cite{Schrittwieser} and are similar to the ones considered in \cite{KillSym}. To be more concrete, we will focus on the following goals: 1.~\textit{Recovery of the control flow graph.} Retrieving the control flow graph of an obfuscated function is a crucial step to understand what the original function performs. 2.~\textit{Detection of opaque predicates}. Recovery of the control flow graph can only be successful if the injected opaque predicates can be detected and removed. 3.~\textit{Deobfuscation of several obfuscation techniques}. To make the code readable and understandable the injected obfuscation patterns have to be detected and removed. 4.~\textit{Recovery of the stack and arguments}. If the attacker can rebuild the stack and arguments, the code of the function will become tidy. 5.~\textit{Execution of the recovered code}. If the attacker is able to execute a semantically equivalent deobfuscated code, further analysis can be done with tools like a debugger if needed.

% Motivating example
\subsection{Motivating example}

\begin{figure}[hbt!]
\centering
\begin{lstlisting}[language=C, 
caption=Anti-symbolic path-oriented protections \textbf{FOR} and \textbf{SPLIT} applied on a toy program based on \cite{KillSym}, label={ToyExample}
]
int func(char chr, char ch1, char ch2) {
  char garb = 0; char ch = 0;
  // FOR trick
  for (int i = 0; i < chr; i++)
    ch++;
  // SPLIT trick
  if (ch1 > 60)
    garb++;
  else
    garb--;
  if (ch2 > 20)
    garb++;
  else
    garb--;
  // MBA based opaque predicate
  if ((chr + ch2) ==  ((chr ^ ch2) + 2 * (chr & ch2)))
    ch ^= 97;
  else
    ch ^= 23;
  return (ch == 31);
}
\end{lstlisting}
\end{figure}

Let us illustrate some anti-symbolic path-oriented protections on a toy program like those introduced in the work of \cite{KillSym}. Listing~\ref{ToyExample} displays an unoptimized obfuscation of a simple toy example that is protected against symbolic execution attacks with the \textbf{FOR} and \textbf{SPLIT} tricks as introduced in \cite{KillSym} and extended with an opaque predicate to protect the final calculation.\\ 

\begin{figure}[hbt!]
\centering
\begin{lstlisting}[language=LLVM, 
caption=Unprotected toy program compiled to \textit{LLVM-IR}, label={ToyExampleIR}
]
define dso_local i32 @func(i8 signext) local_unnamed_addr #0 {
  %2 = icmp eq i8 %0, 126
  %3 = zext i1 %2 to i32
  ret i32 %3
}
\end{lstlisting}
\end{figure}

The anti-symbolic tricks that we considered are not resistant to compiler optimizations and can easily be removed by compiling the code with \textbf{clang -O3} optimization. The introduced opaque predicate is resistant to compiler optimizations and can only be recovered by \textit{SMT} solving. In our tests we will compile the toy example with \textbf{clang -O0} to hinder the optimizer from optimizing away the introduced tricks. The output binary therefore contains several stack slots, that are required to be recovered during the brightening step. If we fail to recover the stack slots and arguments, the \textit{LLVM} optimizations will fail to work and the anti-symbolic tricks won't be removed. If we succeed, the retrieved \textit{LLVM-IR} should look similar to the output of \textbf{clang -O3 -S -emit-llvm} in Listing~\ref{ToyExampleIR} applied on the toy program without any obfuscation.

% Section Function recovery
\section{Function recovery}

Two of the core features of \textit{SATURN} are the exploration and control flow graph reconstruction phases. The \textit{LLVM} ecosystem relies on powerful and correct algorithms that we made use of during the development of \textit{SATURN}'s passes. In this section we explain how \textit{SATURN} achieves full function recovery starting from the binary code.

\subsection{Code lifting to LLVM-IR}

\textit{SATURN} heavily relies on \textit{Remill}. That's why it's important to understand how \textit{Remill} is lifting a native instruction to \textit{LLVM-IR}. \textit{Remill} makes use of the target architecture's CPU instruction semantics to lift an instruction. In Listing~\ref{RemillStruct} we can see the \textbf{State} struct for the \textit{x86\_64} architecture.\\

To emulate an x86\_64 instruction like \textbf{add rax, rcx} \textit{Remill} will create a call to a helper function that implements the emulation for the instruction. This function takes the \textbf{State} variable as an argument (Listing~\ref{RemillFuncDef}) and calculates the result according to the semantic of the instruction. This also includes the \textit{Flags} registers. Once all instructions of a basic block are lifted, the generated calls get inlined into the caller. The output \textit{LLVM-IR} is not very readable at this step, but it behaves functionally the same as the native counterpart.

% Figure for Remill basic block
\begin{figure}[hbt!]
\centering
\begin{lstlisting}[language=C, 
caption=\textit{Remill} basic block function signature C/C++, label=RemillFuncDef
]
Memory *__remill_basic_block(State &state, addr_t curr_pc, Memory *memory);
\end{lstlisting}
\end{figure}

\begin{figure}[hbt!]
 \center
  \includegraphics[width=85mm]{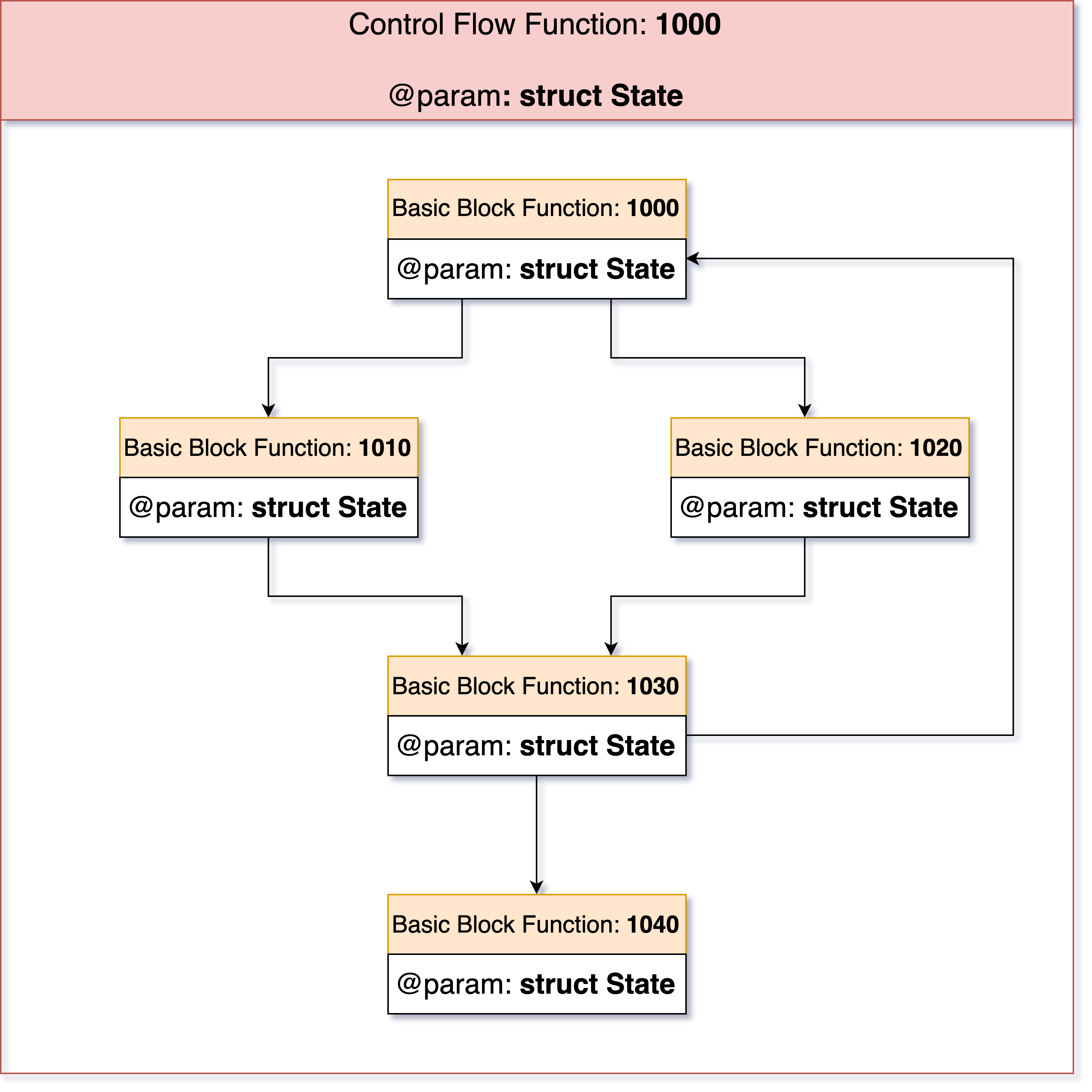}
  \caption{\textit{LLVM} function that contains the control flow graph of the recovered function at address 1000. The basic blocks are lifted as \textit{LLVM} functions themselves and get called according to their usage. The result of the call decides the destination of the branch.}
  \label{fig:CFF}
\end{figure}

During the lifting, \textit{SATURN} stores each recovered basic block into its own \textit{LLVM-IR} function. The basic block functions then get connected in a separate \textit{LLVM-IR} function, which is representing the recovered control flow graph (Figure~\ref{fig:CFF}).\\ 

% Figure for Control Flow Graph function

With this design decision we can directly optimize the lifted basic block function and achieve a performance improvement in further deobfuscation steps. Applying optimizations at this step also removes some simple obfuscation patterns. The control flow function is kept as simple as possible, which allows us to add/remove edges without the need to change the lifted code and avoids dealing with \textit{LLVM's} \textit{PHI}-nodes \cite{Torczon:2007:EC:1526330}.\\

\textit{SATURN} decides how to proceed with the path exploration during the lifting of basic blocks. For that \textit{SATURN} is using the \textit{Remill} instruction categories that are generated for each lifted instruction. Based on the instruction category \textit{SATURN} will try to detect if the basic block is ending with an opaque predicate (Table~\ref{tab:SPE}) or not. If the basic block is a candidate for an opaque predicate, \textit{SATURN} will first try to prove the outgoing edges by applying \textit{LLVM's} optimization passes. If after optimization the count of the outgoing edges is greater than one, \textit{SATURN} will try to solve the outgoing edges by making use of \textit{Souper} and the \textit{Z3} SMT solver \cite{DeMoura:2008:ZES:1792734.1792766}. \textit{SATURN} is always trying to use \textit{LLVM's} optimizations first, since they are much cheaper, performance-wise, compared to the use of an SMT solver. Our tests with various obfuscation engines show that most of the generated opaque predicates are not resistant to \textit{LLVM's} optimizations. The handling of opaque predicates is well described in Section~\ref{chap:DBO}.\\

% Table for Saturn path exploration
\begin{table}[hbt!]
  \caption{Path Exploration}
  \label{tab:SPE}
  \begin{tabular}{lcc}
    \toprule
    kCategory&Exploration&Opaqueness Proof\\
    \midrule
    NoOp&Continue&No\\
    Normal&Continue&No\\
    FunctionReturn&Stop&Yes \\
    IndirectJump&Stop&Yes\\
    DirectJump&Stop&No\\
    ConditionalBranch&Stop&Yes\\
    IndirectFunctionCall&Stop&Yes\\
    DirectFunctionCall&Continue&No\\
  \bottomrule
\end{tabular}
\end{table}

\textit{SATURN} continues with the lifting process as long as new edges are discovered. When \textit{SATURN} is discovering a new incoming edge for a basic block, it has to prove that the new edge does not change the opaqueness of an already (temporarily) proven basic block (Table~\ref{tab:SPE}). The following steps are applied:

\begin{enumerate}
    \item create a new function, called \textbf{FSlice}, based on the definition in Listing~\ref{RemillFuncDef}
    \item find all the basic blocks that dominate the lifted basic block, that we identify as \textbf{BBLift}
    \item if more than one predecessor is found, stop and continue at step 5
    \item repeat step 2 - 3 with the predecessor as input and store the result in a sorted list, called \textbf{PSort}, based on the dominance
    \item for each predecessor in \textbf{PSort} create a call in \textbf{FSlice} in reverse order
    \item connect the called predecessors with a branch instruction
    \item call \textbf{BBLift} at the end of \textbf{FSlice} and connect it to the called dominating predecessor with a branch
    \item inline all calls in \textbf{FSlice} and apply the \textit{LLVM} optimizations.
\end{enumerate}
\hfill

Now \textit{SATURN} is using one of the solutions explained in Section~\ref{chap:BREAKOP} to determine the opaqueness of the basic block.\\

The basic block opaqueness might change to non-opaque and help us to detect false positives. This step is important, as it guides further code exploration. We also need this step because we can't know about all the incoming edges in the beginning and we gain the needed knowledge about the control flow graph only during the exploration phase. The opaqueness of the basic block will change according to Table~\ref{tab:opaqueness}.\\

% Table for opaqueness change
\begin{table}[hbt!]
  \caption{Basic Block Opaqueness}
  \label{tab:opaqueness}
  \begin{tabular}{ccc}
    \toprule
    Current Opaqueness&New Edge&New Opaqueness\\
    \midrule
    No OP&No OP&No OP\\
    No OP&OP&No OP\\
    OP&No OP&No OP\\
    OP&OP&OP\\
  \bottomrule
\end{tabular}
\end{table}

% Section about deobfuscation itself
\section{Deobfuscation by Optimization}
\label{chap:DBO}

The capability to easily build custom optimization passes is part of the core design of \textit{LLVM}. In this section we are going to cover the custom optimization passes that we implemented to facilitate the propagation of constants and the identification of opaque predicates.

\subsection{Constants}
\label{chap:CONSTANTS}

Storing constants in data sections is a common obfuscation technique to trick disassemblers like \textit{IDA Pro} \cite{Eagle:2008:IPB:1481438} into generating wrong results or simply stop the disassembling of the function. During deobfuscation, \textit{SATURN} tries to detect accesses to such constants and replace the \textit{read} instruction with a constant value in the \textit{LLVM-IR}. Demoting the global variables helps the \textit{LLVM} optimization passes to apply constant folding and defeat such kind of obfuscation tricks. The user has to supply the ranges where to look for such constant data with the \textit{SATURN} option \textit{constantPool}. Our tests with several obfuscators show that it's not enough to use the constant binary data sections. In the obfuscators we looked at, we could find sections with read/write attributes that contained such constants.

\subsection{Stack pointer aliasing}
\label{chap:STACKALIASING}

% Solution concrete stack register
\textit{Remill} does not know about the concept of a stack. Instead of trying to emulate the stack, it handles the stack operations by using read and write intrinsics (Listing~\ref{RemillIntrinsicsDef}) relying on the stack register as address. The stack register is part of the \textbf{State} struct and is defined as an \textit{unsigned integer} value like \textbf{uint64\_t State.gpr.rsp.qword} for the \textit{x86\_64} architecture. In \textit{SATURN} the access to the stack will be represented as a load/store of an \textbf{IntToPtr} value. This makes it impossible for \textit{LLVM} to apply pointer aliasing, because \textit{LLVM} does not support pointer aliasing on integer values \cite{LLVMNoAA}.

% Figure for Remill memory read/write intrinsics
\begin{figure}[hbt!]
\centering
\begin{lstlisting}[language=C, 
caption=\textit{Remill}'s memory read and write intrinsics definition, label=RemillIntrinsicsDef
]
uint<T>_t __remill_read_memory_<T>(Memory *, addr_t);

Memory *__remill_write_memory_<T>(Memory *, addr_t, uint<T>_t);
\end{lstlisting}
\end{figure}

In \textit{SATURN} this problem is handled by concretizing the stack register in the function representing the control flow graph. We then inline the basic block functions and optimize the code. During optimization the concrete stack register value will be propagated through the \textit{LLVM-IR} and will replace the \textbf{IntToPtr} operand with a concrete memory location. This concrete value helps us to identify the stack slot. We then create a global variable and a \textit{LLVM-IR} \textbf{Alloca} instruction for the stack slot at the beginning of the control flow graph function. After that we load the value from the global variable and store it into the \textbf{Alloca} value right after the \textbf{Alloca} instruction. We keep a map of known stack slots, their global variables and of the generated \textbf{Alloca} instructions. We optimize the code again and now, based on the \textit{allocas}, \textit{LLVM} is able to apply a proper pointer aliasing pass. These steps may reveal new concrete stack slots and we repeat this algorithm until no new stack slots are detected. Once it's finished we remove the unused global variables.\\

After the algorithm is done, some global variables are not optimized away. These global variables represent the return value, the function arguments passed on the stack and the values popped from the stack and stored in the execution context by the function. This is a side effect that we can use in the further two deobfuscation steps \textit{code brightening} and recovering of the function arguments.\\

Pointer aliasing on the stack is an important feature for deobfuscation.
It's crucial that this step gives accurate results, since it's needed for the following optimization steps.\\

% Fig O3ToyMBA
\begin{figure}[hbt!]
\centering
\begin{lstlisting}[language={[x86masm]Assembler}, caption=Obfuscated x86\_64 \textit{opaque predicate}, label=OB1OP]
0x146253057:  lea   rsp, [rsp-8]
0x14625305F:  push  rcx
0x146253060:  xchg  rcx, [rsp+8]
0x146253065:  mov   rcx, r14
0x146253068:  mov   [rsp+8], rcx
0x14625306D:  mov   rcx, [rsp]
0x146253071:  mov   [rsp], r14
0x146253075:  push  rcx
0x146253076:  lea   rcx, [rsp+8]
0x14625307B:  not   r14
0x14625307E:  xor   r14, [rcx]
0x146253081:  pop   rcx
0x146253082:  push  rbx
0x146253083:  mov   ebx, 0xD4469D6E
0x146253088:  push  rsi
0x146253089:  mov   esi, 0xB7E07B2A
0x14625308E:  add   esi, ebx
0x146253090:  mov   ebx, esi
0x146253092:  xor   ebx, 0x533C089A
0x146253098:  mov   esi, 0xAB832EC0
0x14625309D:  ror   ebx, 0x14
0x1462530A0:  and   esi, 0x5B171CFB
0x1462530A6:  rcl   ebx, 0x1E
0x1462530A9:  or    ebx, 0xE4E97533
0x1462530AF:  shld  esi, ebx, 6
0x1462530B3:  rcl   ebx, 0xD
0x1462530B6:  jb    0x1465C8B69
\end{lstlisting}
\end{figure}

% Subsection Break Opaque Predicates with LLVM-IR Optimizations
\subsection{Breaking Opaque Predicates with LLVM-IR Optimizations}
\label{chap:BREAKOP}

\textit{SATURN} is approaching the opaque predicates problem in two steps. First it creates a slice of the instruction pointer and then applies \textit{LLVM} optimizations on it. If the optimization is successful, the slice will fold into a single concrete value.\\

The available open source slicers (\cite{DBLP}\cite{Chalupa2016thesis}\cite{LLVMSLICER}) seem to be too outdated or unreliable to produce a valid slice for a given function. Conversely, our algorithm is based on modelling the slicing process in \textbf{C} and then relying on the \textit{LLVM} optimizations to produce the slice.

\subsubsection{\textit{SATURN}'s slicing}
\label{chap:SLICINGHELPER}

The \textit{Remill}'s basic block definition in Listing~\ref{RemillFuncDef} contains the information to control and inspect the value of a general purpose register before and after the execution of a \textit{Remill} function. Based on the \textit{Remill} basic block, the slicing is achieved with the following steps:

\begin{enumerate}
\item initialize a \textit{Remill} \textbf{State} struct with a symbolic state
\item concretize the initial instruction pointer (\textbf{RIP})
\item call the opaque basic block, that has been previously optimized with the constant promotion and stack aliasing passes. This call is inlined during further optimization
\item pass the initialized \textbf{State} struct to the basic block to be proven to be opaque
\item get the resulting \textbf{State} struct after the basic block execution, specifically inspecting the final instruction pointer.
\end{enumerate}

% Fig SaturnSlicingHelper
\begin{figure}[hbt!]
\centering
\begin{lstlisting}[language=C, caption=\textit{SATURN}'s slicing helper function, label=SaturnSlicingHelper]
extern "C" uint64_t __saturn_slice_rip(State state, addr_t curr_pc, Memory *memory, uint64_t *Stack) {
	// 1 Allocate a local Remill State structure and initialize it
	State S;	
	S.gpr.rax.qword = state.gpr.rax.qword;
	... 
	S.gpr.rsp.qword = (uint64_t) Stack;
	S.gpr.r15.qword = state.gpr.r15.qword;	
	S.aflag.af = state.aflag.af;	
	...
	S.aflag.zf = state.aflag.zf;
	// 2 Concretize RIP
	S.gpr.rip.qword = curr_pc;
	// 3/4 Call opaque basic block with initialized State struct
	// This function call will be replaced with the lifted one
	__remill_basic_block(S, curr_pc, memory);
	// 5 Inspect the value of RIP
	return S.gpr.rip.qword;
}
\end{lstlisting}
\end{figure}

The initialization and further inspection steps are in Listing~\ref{SaturnSlicingHelper}. The final step is taking the generated \textit{\_\_saturn\_slice\_rip} function and applying \textit{LLVM} optimizations on it. If the function implements an opaque predicate and \textit{LLVM} is able to optimize it away, the function will end with a concrete return value. This is the deterministic instruction pointer address where the basic block will continue. Listing~\ref{OB1OP} is showing an example of the obfuscated opaque predicate. In Listing~\ref{OB1OPTIMIZEDOP} we can appreciate the result of the previously described process. The opaqueness has been broken and the unique destination address recovered.\\

\textit{SATURN} has two options to control the amount of basic blocks to be used while slicing the value of the instruction pointer. This is needed as some obfuscators reuse values from previous basic blocks in subsequent opaque predicates. The \textit{SATURN} options \textit{solverBBCountJcc} and \textit{solverBBCountReturn} let the user specify the amount of basic blocks with a single predecessor to connect to the current opaque block before optimizing it.\\

% Fig 
\begin{figure}[hbt!]
\centering
\begin{lstlisting}[language=LLVM, caption=Sliced and optimized \textit{LLVM-IR} with recovered \textit{Opaque Predicate} offset 0x1465C8B69,
label=OB1OPTIMIZEDOP]
define dso_local i64 @__saturn_slice_rip(%struct.State*, i64, %struct.Memory*, i64*) {
entry:
 ret i64 5475437417	; 0x1465C8B69
}
\end{lstlisting}
\end{figure}

In the next section we approach the problem of hard to optimize opaque predicates, most commonly based on Mixed-Boolean-Arithmetic (\textbf{MBA} \cite{biondi:hal-01241356}) expressions, as seen in our motivating example in Listing~\ref{ToyExample}.

\subsection{Solving Opaque Predicates with Souper and Z3}

The previous approach might fail because \textit{LLVM}'s optimizations are not successful in reducing the sliced instruction pointer to a constant. This means the conditional branch is either based on a stronger opaque predicate or might be a real conditional branch. To further analyze the branch we use the \textit{Souper Optimizer} \cite{Souper} and a \textit{SMT} solver. The steps taken to prove the opaqueness with the \textit{Z3} theorem prover integrated within \textit{Souper} are the following:

% Figure for the sliced MBA Function
\begin{figure}[hbt!]
\centering
\begin{lstlisting}[language=LLVM, 
caption=Sliced MBA of the toy program, label=MBASLICED
]
define i64 @__saturn_slice_rip(%struct.State.32* %state, i64 %curr_pc, %struct.Memory* %memory, i64* %Stack) #2 {
entry:
  %0 = getelementptr inbounds %struct.State.32, %struct.State.32* %state, i64 0, i32 6, i32 17, i32 0, i32 0
  %1 = load i64, i64* %0, align 8, !tbaa !9
  %2 = getelementptr inbounds %struct.State.32, %struct.State.32* %state, i64 0, i32 6, i32 19, i32 0, i32 0
  %3 = load i64, i64* %2, align 8, !tbaa !9
  %4 = shl i64 %1, 56
  %5 = ashr exact i64 %4, 56
  %6 = add i64 %5, %3
  %7 = xor i64 %3, %1
  %8 = shl i64 %7, 56
  %9 = ashr exact i64 %8, 56
  %10 = and i64 %3, %1
  %11 = shl i64 %10, 56
  %12 = ashr exact i64 %11, 55
  %13 = add nsw i64 %12, %9
  %14 = trunc i64 %6 to i32
  %15 = trunc i64 %13 to i32
  %16 = icmp eq i32 %14, %15
  %17 = select i1 %16, i64 5368713261, i64 5368713259
  ret i64 %17
}
\end{lstlisting}
\end{figure}

\begin{enumerate}
\item extract the sliced instruction pointer value from the opaque basic block (Value \%17 in Listing~\ref{MBASLICED})
\item collect a set of candidate expressions to be solved by \textit{Souper}
\item select the \textit{Souper} expression corresponding to the sliced instruction pointer value from the set
\item build an \textit{SMT} query that aims at finding one valid solution for the sliced instruction pointer expression
\item if the query is not satisfiable, something went wrong in the proving process and the pass fails
\item if the query is satisfiable, a valid solution for the expression has been discovered and a second \textit{SMT} query is built to determine if the unique solution has been found, as shown in Listing~\ref{MBASMT}
\item if the last query is satisfiable, the conditional branch has been proven to be opaque and the real destination has been determined
\item if the last query is not satisfiable, a real conditional branch or an opaque predicate which is not provable by the \textit{SMT} solver has been found.
\end{enumerate}

% Figure for the Z3 SMT query for MBA Function
\begin{figure}[hbt!]
\centering
\begin{lstlisting}[language=LLVM, 
caption=\textit{Z3} SMT query to solve the \textbf{MBA} opaque predicate from the toy program in Listing~\ref{ToyExample}, label=MBASMT
]
(set-logic QF_BV )
(declare-fun arr () (_ BitVec 8) )
(declare-fun arr0 () (_ BitVec 8) )
(assert (let ( (?B1 arr0 ) (?B2 arr ) ) (let ( (?B3 ((_ sign_extend 24)  ?B1 ) ) (?B4 ((_ sign_extend 24)  ?B2 ) ) (?B5 (bvand  ?B2 ?B1 ) ) (?B6 (bvxor  ?B2 ?B1 ) ) ) (let ( (?B11 ((_ sign_extend 24)  ?B6 ) ) (?B10 ((_ sign_extend 24)  ?B5 ) ) (?B8 (bvadd  ?B4 ?B3 ) ) (?B9 (bvashr  ?B4 (_ bv31 32) ) ) (?B7 (bvashr  ?B3 (_ bv31 32) ) ) ) (let ( (?B14 ((_ extract 0  0)  ?B9 ) ) (?B12 ((_ extract 0  0)  ?B7 ) ) (?B15 (bvshl  ?B10 (_ bv1 32) ) ) (?B13 (bvashr  ?B8 (_ bv31 32) ) ) (?B16 (bvashr  ?B11 (_ bv31 32) ) ) ) (let ( (?B17 ((_ extract 0  0)  ?B13 ) ) (?B22 ((_ extract 0  0)  ?B16 ) ) (?B19 (bvadd  ?B15 ?B11 ) ) (?B20 (bvashr  ?B15 (_ bv31 32) ) ) (?B21 (bvashr  ?B15 (_ bv1 32) ) ) (?B18 (=  ?B14 ?B12 ) ) ) (let ( (?B27 ((_ extract 0  0)  ?B20 ) ) (?B25 (bvashr  ?B19 (_ bv31 32) ) ) (?B23 (=  ?B17 ?B14 ) ) (?B28 (=  ?B21 ?B10 ) ) (?B24 (=  false ?B18 ) ) (?B26 (=  ?B19 ?B8 ) ) ) (let ( (?B31 (ite  ?B26 (_ bv5368713423 64) (_ bv5368713442 64) ) ) (?B30 ((_ extract 0  0)  ?B25 ) ) (?B29 (or  ?B24 ?B23 ) ) (?B32 (=  ?B27 ?B22 ) ) ) (let ( (?B35 (=  false ?B32 ) ) (?B33 (=  ?B30 ?B27 ) ) (?B34 (=  (_ bv5368713423 64) ?B31 ) ) ) (let ( (?B36 (or  ?B35 ?B33 ) ) ) (let ( (?B37 (and  ?B36 ?B28 ) ) ) (let ( (?B38 (and  ?B37 ?B29 ) ) ) (let ( (?B39 (=  false ?B38 ) ) ) (let ( (?B40 (or  ?B39 ?B34 ) ) ) (and  ?B38 (=  false ?B34 ) ) ) ) ) ) ) ) ) ) ) ) ) ) ) )
(check-sat)
(exit)
\end{lstlisting}
\end{figure}

\section{Recompilation}
\label{chap:Recompilation}

\textit{SATURN} is not only lifting and deobfuscating the code, one of the goals is also to be able to recompile the \textit{LLVM-IR} and make it executable. In this section we explain how \textit{SATURN} is preparing the code for recompilation and how it achieves \textit{vanilla}--like results. 

\subsection{Post Translation Optimization}

Once the obfuscated function is recovered, \textit{SATURN} starts the post translation optimization phase, where the input is the control flow graph function shown in Figure~\ref{fig:CFF}. The steps are as follows:

\begin{enumerate}
\item the stack register (\textbf{RSP}) and the instruction pointer register (\textbf{RIP}) contained in the \textbf{State} variable are concretized
\item \textit{allocas} are created for the flag calculation and stored into the \textbf{State} struct. This helps to optimize and remove unneeded flag calculations
\item the basic block functions are inlined like seen in Figure~\ref{fig:CFF}
\item the \textit{LLVM} optimizations are applied to the function
\item the constant promotion algorithm (Section~\ref{chap:CONSTANTS}) and the stack alias analysis (Section~\ref{chap:STACKALIASING}) are applied to the function
\item the steps 2--4 are repeated until no further changes are detected.
\end{enumerate}

After the post translation is done, the output \textit{LLVM-IR} is in a deobfuscated state but it's still difficult to understand the code because of the operations applied on the \textit{Remill} \textbf{State} struct like shown in Listing~\ref{MBALLSTATE}. At this point the concretization of the registers can be removed and the \textit{LLVM-IR} can be compiled to binary code by making use of one of \textit{LLVM's} backends. In the tests we use \textit{Clang} to compile the output \textit{LLVM-IR} into a shared object. \textit{SATURN} has two options to recompile the \textit{LLVM-IR}:

\begin{itemize}
\item the first option keeps the lifted function with the \textit{Remill} signature as defined in Listing~\ref{RemillFuncDef}. The created \textbf{C++} helper functions do the context switch from the \textit{x86\_64} to the virtual context and take care of the \textbf{State} struct handling;
\item the second option recovers the original function arguments and removes the \textbf{State} struct. This method has the benefit that the function can be called directly without a context switch. This approach is detailed in Section~\ref{chap:CODEBRIGHT}.
\end{itemize}
\hfill

% Section about code brightening and why it's important 
\subsection{Code Brightening}
\label{chap:CODEBRIGHT}

The function lifted by \textit{Remill} is operating on a virtual context (Listing~\ref{MBALLSTATE}), the \textbf{State} struct. This hinders the optimizer from detecting some optimization opportunities, as it has to store the results for each register back to the \textbf{State} struct. This happens for all the registers shown in Listing~\ref{MBALLSTATE}. At this point the output code is still too difficult to understand in further analysis steps like reverse engineering. In this section we address this problem and show how the original signature of the function can be reconstructed. This includes recovering the function arguments and removing the \textbf{State} struct, which leads to \textit{vanilla}--like results.

% Figure for MBA Function before recovering the Arguments
\begin{figure}[hbt!]
\centering
\begin{lstlisting}[language=LLVM, 
caption=Recovered toy program LLVM-IR in the \textit{Remill} \textbf{State} struct form, label=MBALLSTATE
]
define dllexport i64 @F_140001000(%struct.State.32* %S, i64 %curr_pc, %struct.Memory.0* %memory) {
entry:
  %0 = getelementptr inbounds %struct.State.32, %struct.State.32* %S, i64 0, i32 6, i32 33, i32 0, i32 0
  %1 = getelementptr inbounds %struct.State.32, %struct.State.32* %S, i64 0, i32 13
  store i8 0, i8* %1, align 1
  %2 = getelementptr inbounds %struct.State.32, %struct.State.32* %S, i64 0, i32 6, i32 5, i32 0
  %3 = bitcast %union.anon.2* %2 to i8*
  %4 = getelementptr inbounds %struct.State.32, %struct.State.32* %S, i64 0, i32 6, i32 17, i32 0
  %5 = bitcast %union.anon.2* %4 to i8*
  %6 = load i8, i8* %5, align 1
  %7 = load i8, i8* %3, align 1
  store i64 5368713251, i64* %0, align 8
  %8 = sext i8 %7 to i64
  %phitmp = icmp eq i8 %7, 126
  %9 = getelementptr inbounds %struct.State.32, %struct.State.32* %S, i64 0, i32 6, i32 1, i32 0, i32 0
  %10 = getelementptr inbounds %struct.State.32, %struct.State.32* %S, i64 0, i32 6, i32 5, i32 0, i32 0
  %11 = sext i8 %6 to i64
  %12 = and i64 %11, 4294967295
  store i64 5368713372, i64* %0, align 8
  %13 = getelementptr inbounds %struct.State.32, %struct.State.32* %S, i64 0, i32 6, i32 7, i32 0, i32 0
  %14 = xor i8 %7, %6
  %15 = sext i8 %14 to i64
  %16 = getelementptr inbounds %struct.State.32, %struct.State.32* %S, i64 0, i32 6, i32 17, i32 0, i32 0
  store i64 %12, i64* %16, align 8
  %17 = and i64 %12, %8
  %18 = shl nuw nsw i64 %17, 1
  %19 = and i64 %18, 4294967294
  store i64 %19, i64* %13, align 8
  %20 = add nsw i64 %18, %15
  %21 = and i64 %20, 4294967295
  store i64 %21, i64* %10, align 8
  store i8 0, i8* %1, align 1
  %22 = zext i1 %phitmp to i8
  store i8 %22, i8* %3, align 1
  %23 = zext i1 %phitmp to i64
  store i64 %23, i64* %9, align 8
  ret i64 %23
}
\end{lstlisting}
\end{figure}

\subsubsection{Function Arguments}

Based on the algorithm in Section~\ref{chap:STACKALIASING}, the arguments of a lifted function that are passed through the stack, are detected by inspecting the remaining global variables. During the execution of the algorithm in Section~\ref{chap:STACKALIASING}, \textit{SATURN} keeps track of the global variables and their stack offsets. This information will be used to detect the number of arguments, with the knowledge about the application binary interface (\textbf{ABI}) and the calling convention \cite{MSABI} used by the function. If no stack arguments are passed to the function, we detect the number of arguments through the register accesses on the \textbf{State} struct. We only focus on the reconstruction of the general purpose registers in the following steps:

\begin{enumerate}
\item based on the function's \textit{calling convention}, start with the last register argument in the function's \cite{MSABI} argument list and search for the first \textit{getElementPtr} (\textbf{GEP}) instruction that's accessing the register and is also dominating all the other \textbf{GEP} instructions that access that register
\item if no \textbf{GEP} instruction is found, continue with the next register and decrease the number of arguments by one
\item if a \textbf{GEP} instruction was found, forward slice the \textbf{GEP} value to get a tree of users that have a reference to the \textbf{GEP} instruction
\item sort the users based on their position in the dominance tree \textbf{DT} of the function
\item look for \textbf{load} and \textbf{store} instructions to detect how the \textbf{GEP} is used
\item if a dominating \textbf{load} or \textbf{store} can be found, assume that this register is an argument
\item else decrease the number of arguments by one and continue at step 3.
\end{enumerate}
\hfill

\subsubsection{Function reconstruction}

Based on the recovered number of arguments we start to rebuild the lifted function to be detached from the \textbf{State} struct. We use helper functions in \textbf{C/C++} that assist us to easily map the function arguments to their slots in the \textbf{State} struct as shown in Listing~\ref{SaturnArgsHelper}.

% Figure for the SATURN argument helper function
\begin{figure}[hbt!]
\centering
\begin{lstlisting}[language=C, 
caption=\textit{SATURN}'s helper C/C++ function to handle a Windows 64-bit ABI function with 2 arguments, label=SaturnArgsHelper
]
extern "C" Memory * F_Lifted(State &state, addr_t curr_pc, Memory *memory);
extern "C" uint64_t x64_MS_2_ARG(uint64_t *RCX, uint64_t *RDX) {
	struct State S;
	// Set 1. arg
	S.gpr.rcx.qword = (uint64_t) RCX;	
	// Set 2. arg
	S.gpr.rdx.qword = (uint64_t) RDX;	
	// Call lifted function which will be replaced and inlined	
	F_Lifted(S, 0, nullptr);
	// Return result
	return S.gpr.rax.qword;
}
\end{lstlisting}
\end{figure}

We only need to prepare helpers for the register based arguments. On functions that pass arguments on the stack we can simply add new arguments to the helper function in the \textit{LLVM-IR} and replace all the references of the global value representing the stack argument to the newly created function argument. The further steps are independent from the number of arguments:

\begin{enumerate}
\item find the call to the \textbf{F\_Lifted} dummy function
\item replace the reference of \textbf{F\_Lifted} to the lifted function
\item inline the call into the helper function
\item run \textit{LLVM}'s strongest optimizations.
\end{enumerate}
\hfill

Based on \textit{LLVM}'s optimizations we get a clean \textit{LLVM-IR} function that looks \textit{vanilla}--like as shown in Listing~\ref{MBARECOVERED}. If we compare the input \textit{LLVM-IR} in Listing~\ref{MBALLSTATE} and the result in Listing~\ref{MBARECOVERED}, we can see how strong and effective the \textit{LLVM} optimizations are.

% Figure for MBA Function with recovered arguments
\begin{figure}[hbt!]
\centering
\begin{lstlisting}[language=LLVM, 
caption=Optimized \textbf{MBA} LLVM-IR function with recovered arguments, label=MBARECOVERED
]
define dllexport i64 @F_140001000_args(i64* %RCX, i64* %RDX, i64* %R8) {
entry:
  %0 = ptrtoint i64* %RCX to i64
  %1 = trunc i64 %0 to i8
  %2 = icmp eq i8 %1, 126
  %3 = zext i1 %2 to i64
  ret i64 %3
}
\end{lstlisting}
\end{figure}

% Section about execution of the recovered code 
\section{Execution}

\textit{SATURN} is not only able to lift, deobfuscate and brighten the code. It's also able to inject the deobfuscated function back into the input binary. Based on the recovery result (with or without \textbf{State} struct) there are two different ways to call the recovered function.\\

In both described ways the shared library gets injected into the input binary. In portable executable (\textbf{PE}) files the \textit{import table} gets replaced with an updated \textit{import table} that contains an \textit{import} to a function in our shared library from Section~\ref{chap:Recompilation}.

\subsection{Direct Function Redirection}

When \textit{SATURN} is able to fully recover the function and its arguments, we can choose to patch the original function and insert a branch instruction to the imported symbol.

\subsection{Context Switch}

If the recovery of the function arguments fails, \textit{SATURN} is able to keep the \textbf{State} struct in the recovered function. This approach needs a more advanced way to execute the function. The needed \textit{runtime} is implemented in \textbf{C++} and \textbf{x86\_64 assembly}. The \textit{runtime} will be compiled into the shared library that is generated in Section~\ref{chap:Recompilation}. The needed steps to call the function are:

\begin{enumerate}
\item patch an instruction at the beginning of the obfuscated function to push one integer value on the stack (used as function identifier)
\item patch a second instruction to jump into our imported symbol in the import table.
\end{enumerate}

When the function is reached during execution, it will jump into our \textit{runtime} that does the following:

\begin{enumerate}
\item create a virtual stack and use it in place of the original one
\item store all the register values into a local \textbf{State} struct
\item call the lifted function with the generated \textbf{State} struct
\item on calls/jumps outside of the lifted function restore the registers from the \textbf{State} struct and handle the return of the function
\item if the function returns, restore the registers and jump back to the caller.
\end{enumerate}
\hfill

% Experimental evaluation
% \footnote{\label{t1}Remill \textbf{State} struct was used}

% Table Dataset Results
\begin{table*}[hbt!]
\small
\caption{Results for datasets}
\centering
\begin{tabular}{ |p{1,3cm}||p{0,8cm}|p{0,9cm}|p{1,1cm}|p{0,8cm}|p{1,3cm}|p{0,9cm}|p{1,1cm}|p{1,1cm}|p{1,4cm}|p{1,5cm}|p{1,1cm}|  }
 \hline
 \multicolumn{12}{|c|}{\textbf{Dataset \#1}} \\
 \hline
 \centering Program & \centering Time to lift & \centering Time to optimize & \centering Detected Opaque Predicates & \centering Test passed & \centering Arguments recovered (Lift./Orig.) & \centering Stack Slots recovered & \centering Processed Instructions & \centering Recovered Basic Blocks & \centering Obfuscation removed & \centering Solving time with KLEE (Lift./Orig.) & Size reduction in Basic Blocks\\
 \hline
 args               & \centering 0.541s  & \centering 0.980s & \centering 0/0$^{a}$ & \centering Yes$^{a}$ & \centering 0/6$^{a}$ & \centering 0$^{a}$ & \centering 110 & \centering 13/13 & Yes & \centering  - & 4 \\
 cmp\_test          & \centering 0.408s  & \centering 0.342s & \centering 1/1 & \centering Yes & \centering 2/2 & \centering 7 & \centering 39 & \centering 8/8 & Yes & \centering - & 5\\
 edges              & \centering 0.129s  & \centering 0.233s & \centering 2/2 & \centering Yes & \centering 1/1 & \centering 5 & \centering 16 & \centering 4/4 & Yes & \centering - & 3\\
 edges2             & \centering 0.428s  & \centering 0.622s & \centering 0/0 & \centering Yes & \centering 2/3$^{d}$ & \centering 11 & \centering 120 & \centering 10/10 & Yes & \centering - & 9\\
 fib                & \centering 0.309s  & \centering 0.287s & \centering 0/0 & \centering Yes & \centering 1/1 & \centering 6 & \centering 31 & \centering 4/4 & Yes & \centering - & 0\\
 gotos              & \centering 0.599s  & \centering 0.564s & \centering 3/3 & \centering Yes & \centering 3/3 & \centering 10 & \centering 85 & \centering 13/13 & Yes & \centering - & 10\\
 inf\_loop          & \centering 0.215s  & \centering 0.285s & \centering 1/1 & \centering Yes & \centering 1/1 & \centering 0 & \centering 18 & \centering 2/2 & Yes & \centering - & 0\\
 loop               & \centering 0.152s  & \centering 0.247s & \centering 0/0 & \centering Yes & \centering 1/1 & \centering 5 & \centering 21 & \centering 4/4 & Yes & \centering - & 3\\
 multiedges         & \centering 0.405s  & \centering 0.251s & \centering 0/0 & \centering Yes & \centering 1/1 & \centering 4 & \centering 21 & \centering 9/9 & Yes & \centering - & 8\\
 op1                & \centering  0.188s  & \centering 0.010s & \centering 1/1 & \centering Yes & \centering 2/2 & \centering 5 & \centering 24 & \centering 3/3 & Partially$^{e}$ & \centering - & 2\\
 tig\_virt          & \centering 2.337s  & \centering 1.018s & \centering 0/0 & \centering Yes & \centering 1/1 & \centering 17 & \centering 288 & \centering 41/41 & No & \centering - & 23\\
 sse2          & \centering 0.147s  & \centering 0.429s & \centering 0/0 & \centering Yes & \centering 2/2 & \centering 16 & \centering 47 & \centering 1/1 & Yes$^{f}$ & \centering - & 0\\
 \hline
 \multicolumn{12}{|c|}{\textbf{Dataset \#2}} \\
 \hline
 binsec0            & \centering 4.785s  & \centering 1.144s & \centering 0/0 & \centering Yes & \centering 1/1 & \centering 19 & \centering 226 & \centering 25/25 & Yes & \centering 0.43s/2m6.30s & 9\\
 binsec0\_virt      & \centering 6.062s  & \centering 0.595s & \centering 0/0 & \centering Yes$^{a}$ & \centering 0/2 & \centering 0$^{a}$ & \centering 464 & \centering 86/86 & No & \centering - & 36\\
 binsec1            & \centering 4.778s  & \centering 0.141s & \centering 0/0 & \centering Yes$^{a}$ & \centering 0/2 & \centering 0$^{a}$ & \centering 273 & \centering 27/27 & Yes & \centering - & 3\\
 forsplit           & \centering 3.033s  & \centering 0.738s & \centering 0/0 & \centering Yes & \centering 3/3 & \centering 8 & \centering 112 & \centering 12/12 & Yes & \centering 0.08s/16.13s & 4\\ 
 \hline
 \multicolumn{12}{|c|}{\textbf{Dataset \#3}} \\
 \hline
 obf0\_virt         & \centering 27.998s & \centering 8.592s & \centering 0/0 & \centering No$^{b}$ & \centering 4/? & \centering 27 & \centering 2059 & \centering 10/?$^{c}$ & Yes & \centering - & 4\\
 obf1\_func0        & \centering 0.754s & \centering 0.680s & \centering 11/11 & \centering No$^{b}$ & \centering 0/0 & \centering 10 & \centering 182 & \centering 12/12 & Yes & \centering - & 10\\
 obf1\_func1        & \centering 1.060s & \centering 0.827s & \centering 0/0 & \centering No$^{b}$ & \centering 0/0 & \centering 9 & \centering 241 & \centering 14/14 & Yes & \centering - & 13\\
 \hline
 \multicolumn{12}{l}{$^{a}$Remill \textbf{State} struct was used $^{b}$Binary execution failed because of other protections like anti-tampering} \\
 \multicolumn{12}{l}{$^{c}$Unknown amount of original basic blocks $^{d}$Dead argument was detected $^{e}$MBA formula was not optimized away $^{f}$Recovered integer arithmetic}\\
\end{tabular}
\label{Fig:DS}
\end{table*}

\section{Experimental evaluation}
The experiments below seek to answer the following Research Questions:

\begin{enumerate}[label=\subscript{\textbf{RQ}}{{\textbf{\arabic*}}}]
    \item
    What is the effectiveness on the recovery of the control flow graph?
    \item
    What is the detection rate of the opaque predicates?
    \item 
    What is the effectiveness of the deobfuscation?
    \item
    Were all arguments and stack slots recovered?
    \item 
    Is the deobfuscated code semantically equivalent to the protected one during execution?
\end{enumerate}

% Experimental setup
\subsection{Experimental setup}
The attacker has access to \textit{SATURN} to reverse engineer the given binaries and has the goal to recover the obfuscated functions. For the defense we created some binaries that trigger corner cases and use several obfuscation techniques that might hinder binary lifting. Some binaries are protected by Tigress \cite{Tigress}, an open source \textit{state--of--the--art} obfuscator. Some other binaries are protected by real world obfuscators where no source code is available. The machine for the experiments uses Windows 10 Pro x64 on a Intel Core i7--6700k CPU with 32 GB RAM.

% Dataset
\subsection{Datasets}

We select some small programs that trigger corner cases in several steps of the approach described in this paper. We also choose programs that contain selected obfuscation patterns and real world obfuscated binaries where no source code is available. The datasets and results are available in our online repository \footnote{https://github.com/pgarba/Saturn\_Results}.\\

\textbf{Dataset \#1}. During the development of \textit{SATURN} we created several test samples that can trigger corner cases. The tests include scenarios like overlapping stack slots, register and stack based arguments, loops, infinite loops, opaque predicates, MBA based opaque predicates and dead code. The programs take an input value from the command line, perform some calculations and print the output to the user. The sample \textit{tigress\_virtualize} is protected with the \textit{tigress} virtualization obfuscation pass (\textit{--Transform=Virtualize}).\\

\textbf{Dataset \#2}. This dataset includes some programs that were taken from the repository provided by \cite{KillSym} and use the \textit{SPLIT} and \textit{FOR} anti--\textit{DSE} tricks.\\

\textbf{Dataset \#3}. The last dataset contains two real world binaries that were protected with \textit{Obfuscator0} and \textit{Obfuscator1} \footnote{\textit{Obfuscator0} and \textit{Obfuscator1} are made up names}. The binary from \textit{Obfuscator1} was chosen because we have an unprotected binary and can easily compare the results. The  \textit{Obfuscator0} doesn't have an associated unprotected binary, but we choose it because it's a strong obfuscated real world example of a protected virtual machine entry point\footnote{Context switch from the original \textit{x86\_64} to the virtual context}. Both binaries are heavily obfuscated and were chosen to stress test \textit{SATURN}.

% Present the results and match it to the goals
\subsection{Results \& Observations}

Table~\ref{Fig:DS} shows the results for each tested program in the dataset. In programs for which the argument and stack recovery fails $^{a}$, we are still able to recover the function by staying in the Remill \textbf{State} struct (\textbf{RQ1}). For all other programs the recovery of the arguments and the stack was successful (\textbf{RQ4}). In the programs \textit{obf0\_func0} and \textit{obf1\_func0} we don't know the exact amount of opaque predicates but based on the obfuscator we know that a missed opaque predicate would lead to a broken \textit{LLVM-IR}. For the programs in dataset \#1 and \#2 all opaque predicates are detected (\textbf{RQ2}).\\

For each program we verified the deobfuscated function for correctness by comparing the output values to the one of the obfuscated program. Programs protected with the \textit{Tigress} virtualization stay in the virtualized form but the recovered code is clean and readable. The program \textit{op1} is only partially deobfuscated. The \textit{MBA} based opaque predicate is removed but the recovered calculation is still based on the \textit{MBA} formula $^{e}$. The result of \textit{obf0\_virt} can't be verified but the recovered code is clean, readable and meaningful (\textbf{RQ3}). For the programs in dataset \#2 we are not able to remove the \textit{FOR} and \textit{SPLIT} tricks in the Tigress protected samples. In the other programs the tricks are detected and removed.\\

For dataset \#1 and \#2 we are able to execute the output binary and all the deobfuscated programs behave in a semantically equivalent way to the obfuscated ones (\textbf{RQ5}). For dataset \#3 the recompiled binaries are not working because of some additional anti--tampering checks in those binaries $^{b}$.\\

For all the programs we verified the output binary obtained by \textit{SATURN} with the \textit{IDA Pro} \cite{Eagle:2008:IPB:1481438} decompiler. The decompiler is returning meaningful and readable pseudo \textbf{C} code (\textbf{RQ3}). This was failing before as \textit{IDA Pro} struggles with obfuscated code.

% Discussion
\section{Discussion}

% Compare to other binary lifters
We compared our work to existing \textit{LLVM} based binary lifting frameworks. All of them were failing in lifting obfuscated code (\cite {Yadavalli:2019:RBL:3316482.3326354}, \cite{RETDEC_BOTCONF_17}, \cite{MCSEMA}, \cite{Korencik2019thesis}) as they were not built for this task. A good overview of the existing \textit{LLVM}--based lifters is given in the comparison table that can be found in \cite{MCSEMA}. One exception is \textit{S2E} \cite{Chipounov:2012:SPD:2110356.2110358}, the symbolic execution tool based on \textit{QEMU} \cite{Bellard:2005:QFP:1247360.1247401}. \textit{S2E} is able to export the generated traces in pure \textit{LLVM-IR} form but, considering it is a dynamic approach, we can't compare it to our work.\\

% Triton: say that we also tried with triton for op solving but our method is superior because we optimize the llvm ir and get much smaller SMT queries compared to the triton one which results in a much better performance
% Triton
We also compared our work to the symbolic execution tool \textit{Triton} \cite{Tillet:2019:TIL:3315508.3329973}. We were interested to see how \textit{SATURN} compares to \textit{Triton} while processing the opaque predicates with an \textit{SMT} solver. We noticed that \textit{SATURN} is able to create much smaller and optimized \textit{SMT} queries due to prior optimizations. In this regard our approach is much more efficient compared to the one in \textit{Triton} and therefore reduces the solving times\footnote{The comparison is available in the results' repository}. This complies with the assumption in \cite{biondi:hal-01241356}.\\

% TODO maybe add something about LOOP here
The work presented in \cite{Ming:2015:LLO:2810103.2813617}, although based on binary execution traces, is a valid starting point to improve the detection of the dynamic opaque predicates in \textit{SATURN}. While the work presented in \cite{biondi:hal-01241356} describes a strong simplification methodology based on the \textit{Drill\&Join} synthesis technique \cite{10.1007/978-3-319-17822-6_13} which is orthogonal to the ones in \textit{SATURN} and could further improve the MBA expressions handling. As discussed in Section~\ref{chap:FW}, a plugin system would enable us to integrate these approaches during the exploration phase.

% Related Work
\section{Related Work}

% Souper database and opaque predicates
\textbf{Machine Learning.} One of the side products of \textit{SATURN} is a database with normalized opaque predicates and obfuscation patterns thanks to the \textit{Redis} \cite{Redis} cache used in the \textit{Souper Optimizer} \cite{Souper}. We think that it would be interesting to see what a machine learning based method like the one introduced in the work of Tofighi-Shirazi et al. \cite{tofighishirazi:hal-02269192} could make of this information to improve the opaque predicate recovery rate in \textit{SATURN}.\\

% SMT related work like gadget create AEG ...
\textbf{Exploit generation.} 
Rolles et al. described how \textit{SMT} solving can help to automatically chain together sequences of \textit{ROP} gadgets, so that the sequence is semantically equivalent to a model payload \cite{SMTRolf}. We think that \textit{SATURN} can be used to create such \textit{ROP} gadgets and, in combination with a \textit{DSE} tool like \textit{KLEE} \cite{KLEE} model the needed sequence and payload for an exploit.\\

% AutomaticDeobfuscation asked for a tool like this to minimize/optimize the SMT formula 
% Use this so solve MBAs
\textbf{Effectiveness of Synthesis in Concolic Deobfuscation.} Biondi et al. summarize in \cite{biondi:hal-01241356} that \textit{SMT} solvers alone are not efficient enough against \textit{MBA} based obfuscation. As future work they proposed to study a tool that could drive the concolic execution of obfuscated programs by retrieving a compact representation of obfuscated constraints. They expect that the simplified constraints are easier to solve or check for satisfiability. We believe that \textit{SATURN} is exactly the tool that they are looking for. It would be interesting to study the work done in \textit{SATURN} in combination with the work in \cite{biondi:hal-01241356}.

% Conclusion
\section{Conclusion}

In this paper we have proposed a \textit{state--of--the--art} framework for software deobfuscation based around the \textit{LLVM} ecosystem. The work implemented in the tool \textit{SATURN} lifts the attack surface away from the binary level up to the \textit{LLVM-IR} and solves the problems that appear during binary deobfuscation directly on this level. The results that we reach are not based on any assumptions, instead we use general optimization techniques and \textit{SMT} solving to extract the control flow graph and deobfuscate the code. The achieved optimized representation can help to apply advanced practical attacks. We believe that the presented work highlights a new perspective on program deobfuscation and complements existing work by lifting the attack surface to a new level.

% Future work
\section{Future Work}
\label{chap:FW}

We would like to add a plugin system to let the user hook in several phases of the code recovery in \textit{SATURN} and write their own transformation passes. This could be used to devirtualize a protection like the \textit{Tigress} virtualization or handle \textit{MBA} expressions with customized approaches. Right now we are concretizing the stack pointer to be able to retrieve the stack slots, but we think that we could change this step to be based on a completely symbolic approach. We would also like to try out some new ideas in which we change the type of the registers used in \textit{Remill} into pointers, as this may help to avoid \textit{IntToPtr} casts in \textit{LLVM} and generate cleaner code to begin with. Right now we are only able to lift \textit{x86\_64} binary code but \textit{Remill} also supports lifting of \textit{AArch64} and \textit{x86} binary code. We are aware that a constant range analysis was recently added to the constant synthesis in \textit{Souper} \cite{SouperRepo}. We believe this could be used to tackle the difficulties related with the identification of switch-case destination addresses.
 
%% acknowledgments
\begin{acks}
To Roman, for taking the time to proofread our work and also contributing with his knowledge during countless discussions.
To Helge, who cared for guiding me during many years of my career as a reverse engineer.
To Davide, Joao and Massimo for introducing me to the software protections world and for the endless discussions about reverse engineering.
\end{acks}

\bibliographystyle{ACM-Reference-Format}
\bibliography{sample-base}

\end{document}